\documentclass[
reprint,
superscriptaddress,
 amsmath,
 amssymb,
 aps,
prl,
]{revtex4-2}

\usepackage{graphicx}% Include figure files
\usepackage{epstopdf}
\usepackage{dcolumn}% Align table columns on decimal point
\usepackage{bm}% bold math
\usepackage{todonotes} %Adds todo commands
\usepackage{physics}
\usepackage{siunitx}
\usepackage[colorlinks=true, linkcolor=blue, citecolor=blue, urlcolor=blue]{hyperref}
\usepackage{siunitx}
\usepackage{braket}

\usepackage{color}
\usepackage{colordvi}  
\definecolor{Red}{rgb}{1,0,0}

\newcommand{\q}{\mathbf{q}}

\makeatletter
\def\authornote{\xdef\@thefnmark{$\dagger$}\@footnotetext}
\makeatother

\begin{document}

\title{Signatures of dynamically dressed states}
\author{Katarina Boos$^\dagger$}
 \affiliation{Walter Schottky Institut, TUM School of Computation, Information and Technology, and MCQST, Technische Universit\"at M\"unchen, 85748 Garching, Germany}%
\author{Sang Kyu Kim$^\dagger$}
 \affiliation{Walter Schottky Institut, TUM School of Computation, Information and Technology, and MCQST, Technische Universit\"at M\"unchen, 85748 Garching, Germany}%
 \authornote{These authors contributed equally.}
\author{Thomas Bracht} 
\affiliation{Condensed Matter Theory, TU Dortmund, 44221 Dortmund, Germany}%
\affiliation{Institut für Festkörpertheorie, Universität Münster, 48149 Münster, Germany}
\author{Friedrich Sbresny}
 \affiliation{Walter Schottky Institut, TUM School of Computation, Information and Technology, and MCQST, Technische Universit\"at M\"unchen, 85748 Garching, Germany}%
 \author{Jan Kaspari}
 \affiliation{Condensed Matter Theory, TU Dortmund, 44221 Dortmund, Germany}%
 \author{Moritz Cygorek}
 \affiliation{Institute of Photonics and Quantum Sciences, Heriot-Watt University, Edinburgh EH14 4AS, United Kingdom}%
  \author{Hubert Riedl}
 \affiliation{Walter Schottky Institut, TUM School of Natural Sciences, and MCQST, Technische Universit\"at M\"unchen, 85748 Garching, Germany}%
 \author{Frederik W. Bopp}
 \affiliation{Walter Schottky Institut, TUM School of Natural Sciences, and MCQST, Technische Universit\"at M\"unchen, 85748 Garching, Germany}%
 \author{William Rauhaus}
 \affiliation{Walter Schottky Institut, TUM School of Computation, Information and Technology, and MCQST, Technische Universit\"at M\"unchen, 85748 Garching, Germany}%
 \author{Carolin Calcagno}
 \affiliation{Walter Schottky Institut, TUM School of Computation, Information and Technology, and MCQST, Technische Universit\"at M\"unchen, 85748 Garching, Germany}%
\author{Jonathan J. Finley}
 \affiliation{Walter Schottky Institut, TUM School of Natural Sciences, and MCQST, Technische Universit\"at M\"unchen, 85748 Garching, Germany}%
\author{Doris E. Reiter}
 \affiliation{Condensed Matter Theory, TU Dortmund, 44221 Dortmund, Germany}%
\author{Kai M\"uller}
 \affiliation{Walter Schottky Institut, TUM School of Computation, Information and Technology, and MCQST, Technische Universit\"at M\"unchen, 85748 Garching, Germany}%

\date{\today}

\begin{abstract}
The interaction of a resonant light field with a quantum two-level system is of key interest both for fundamental quantum optics and quantum technological applications employing resonant excitation. While emission under resonant continuous-wave excitation has been well-studied, the more complex emission spectrum of dynamically dressed states – a quantum two-level system driven by resonant pulsed excitation - has so far been investigated in detail only theoretically. Here, we present the first experimental observation of the complete resonance fluorescence emission spectrum of a single quantum two-level system, in form of an excitonic transition in a semiconductor quantum dot, driven by finite Gaussian pulses. We observe multiple emerging sidebands as predicted by theory with an increase of their number and spectral detuning with excitation pulse intensity and a dependence of their spectral shape and intensity on the pulse length. Detuning-dependent measurements provide additional insights into the emission features. The experimental results are in excellent agreement with theoretical calculations of the emission spectra, corroborating our findings.
\end{abstract}

\pacs{Valid PACS appear here}

\maketitle

% introduction
Resonance fluorescence, a quantum two-level system driven by a resonant light field, manifests the rich nature of coherent light-matter interaction~\cite{muller2007, vamivakas2009, Matthiesen2012, Prechtel2013, tomm2023}. It has indeed been one of the central topics of quantum optics since the early stages of quantum mechanics~\cite{mollow1969, dirac1927, weisskopf1931, Heitler1936}, interesting for both fundamental research and applications in quantum technologies~\cite{carreno2018, senellart2017}. To study the phenomena in experiment, several different systems have been employed, with quantum dots (QDs) standing out due to their excellent optical properties and their strong light-matter interaction~\cite{senellart2017, Gazzano2013, Schweickert2018, Tomm2021}. Excitonic transitions in QDs behave as prototypical quantum two-level systems which are optically accessible and controllable~\cite{zrenner2002, Ramsay2010, Reiter2014, Zhou2023}. While the fluorescence spectra under resonant continuous-wave excitation have been theoretically and experimentally well-studied~\cite{mollow1969, vamivakas2009, ulhaq2012, Ulhaq2013, Reiter2017, gustin2021,  Ramsay2010_2}, interaction and emission dynamics of QDs driven by finite pulses have been significantly less explored. This is surprising, as resonant pulses are commonly used to drive transitions in QDs~\cite{press2008, stievater2001, zrenner2002, Ramsay2010_2} and prepare them in specific states for subsequent photon emission, i.e. in applications as on-demand single-photon sources~\cite{senellart2017, Zhou2023}. However, time-dependently dressed states in pulsed resonant excitation leave trace of their coherent dynamics in the appearance of multiple peaks in fluorescence spectra which have been so far been only theoretically studied in atoms~\cite{rzazewski1984, florjaczyk1985, lewenstein1986, buffa1988, rodgers1991} and QDs~\cite{Moelbjerg2012, gustin2018}, experimentally observed for ensembles in microplasma~\cite{compton2009, compton2011} and in a QD cavity system~\cite{Fischer2016} where the cavity however significantly modifies the dynamics and masks key features of the rich emission spectrum.

\begin{figure}
    \centering
    \includegraphics[width=\linewidth]{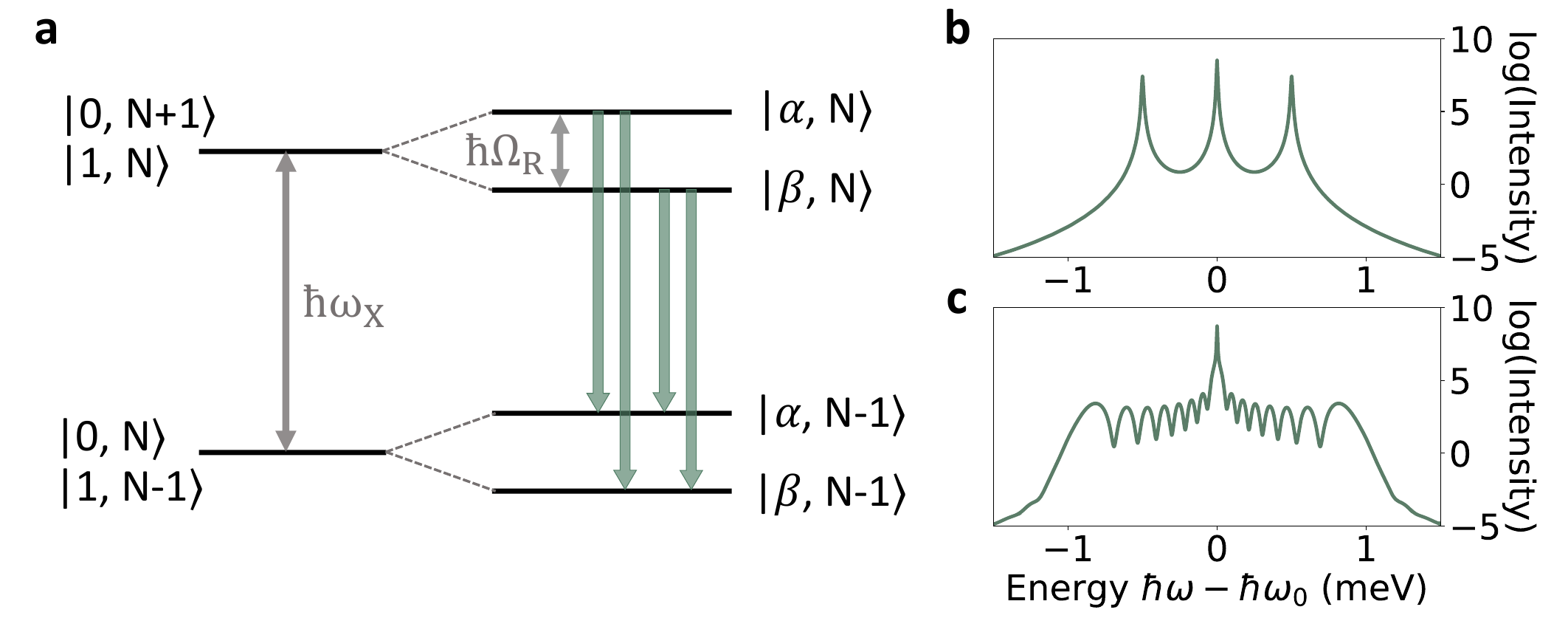}
    \caption{Sketch of dressed states and simulated emission spectra for resonant constant and dynamic dressing. \textbf{a}, For a given intensity of the excitation laser the degenerate levels split up forming new eigenstates. \textbf{b}, While this results in a Mollow triplet in emission for continuous-wave excitation, \textbf{c} excitation with finite Gaussian pulses lead to an emission spectrum with multiple sidepeaks.}
    \label{fig:graph1}
\end{figure}

In this letter, we investigate the full emission spectrum of a semiconductor QD under resonant and near-resonant excitation with a finite Gaussian laser pulse. As theoretically proposed by Moelbjerg \textit{et al.} in 2012~\cite{Moelbjerg2012}, we observe with increasing Rabi energy of the excitation multiple sidepeaks emerging and linearly shifting outward from the transition. To analyze the characteristics of these sidepeaks, we study the impact of the pulse duration on the emission spectra and the behaviour of the sidebands under finite detuning of the excitation from resonance. Theoretical calculations of the emission spectra are in excellent agreement with the experimental data and confirm our findings.\\
\\
% theory and concept
When a quantum two-level system is perturbed by an external laser field, the bare states of the system together with the laser form so-called dressed states.
The energy and composition of the dressed states depend on amplitude and relative detuning of the laser compared to the quantum two-level system (Fig.~\ref{fig:graph1}a). In the resonance fluorescence spectrum of a quantum emitter under continuous-wave excitation, emission from the dressed states is visible as the well-known Mollow-triplet, as depicted in Fig.~\ref{fig:graph1}b, where $\omega_0$ is the transition frequency of the two-level system. Here, the Rabi frequency of the laser, given by $\Omega_{R}=\mu E_0 / \hbar$, with electric field amplitude $E_0$ and electric dipole moment of the QD $\mu$, results in peaks of the triplet visible at $\omega_0 \pm \Omega_{R}$ where $\hbar\Omega_{R}$ is the energetic splitting of the dressed states. 
For excitation with finite Gaussian pulses, dressing of the states persists only on the timescale of the presence of the pulse which is short relative to the excited state lifetime, and the strength of the dressing is modulated by the Gaussian shape of the pulse. This leads to a more complex emission spectrum as shown in the simulated spectrum in Fig.~\ref{fig:graph1}c, where interference of emission at different times during the presence of the pulse gives rise to multiple sidepeaks in the time-integrated spectrum~\cite{compton2011, Moelbjerg2012}.\\
To calculate the emission spectrum of the QD, we calculate the two-time correlation function $G^{(1)}(t,\tau) = \braket{\sigma^\dagger(t+\tau)\sigma(t)}$, where $\sigma = \ket{g}\bra{x}$ is the transition operator of the two-level system. To this end, we apply the techniques presented in Ref.~\cite{cosacchi2018,cygorek2022}. The spectrum of the pulsed excitation is then retrieved from the correlation function as presented in Ref.~\cite{Moelbjerg2012,cosacchi2018} (also see supplemental material~\cite{Supplemental}). Most importantly, the numerical exact treatment of the electron-phonon interaction~\cite{luker2019} using the path-integral formalism~\cite{vagov2011,barth2016,cygorek2017,cygorek2022} allows for calculating the multi-time correlation functions without the application of the quantum regression theorem, which introduces systematical errors if the system is influenced by non-Markovian effects like the electron-phonon interaction \cite{cosacchi2021}.

% experiment and sample
To experimentally investigate the characteristics of the pulsed resonance fluorescence spectrum, we studied the negative trion transition $X^-$- $e^-$ of a single InGaAs QD as a two-level system. The QD is embedded in a Schottky diode to deterministically control the charge state and stabilize the electronic environment~\cite{warburton2000, seidl2005}. A distributed Bragg reflector situated beneath the QD forms a weak planar resonator together with the top interface and enhances light-matter coupling and collection of the photon emission. 
The sample is cooled down to $\SI{4.2}{K}$ in a custom-made helium dip stick, where the excitation and detection channels of this resonance fluorescence setup are separated by cross-polarization filtering~\cite{vamivakas2009}. To generate excitation pulses with variable duration and a Gaussian shape, we shape the $\SI{150}{\femto\second}$ pulse of a Ti:Sapph laser employing a folded 4f pulseshaper~\cite{Weiner1988}. For each recorded spectrum in the experiment, we averaged between 5 spectra (integration time $\SI{2}{\second}$) at resonance and subtracted 5 averaged spectra where the QD is electrically tuned out of resonance to account for background and non-perfect laser suppression. Electrical readout noise in combination with this measurement procedure leads in part to negative values, thus an overall offset of 10 counts is added to the spectra to enable reasonable presentation of the data in colormaps with arbitrary scaling.

\begin{figure}
    \centering
    \includegraphics[width=\linewidth]{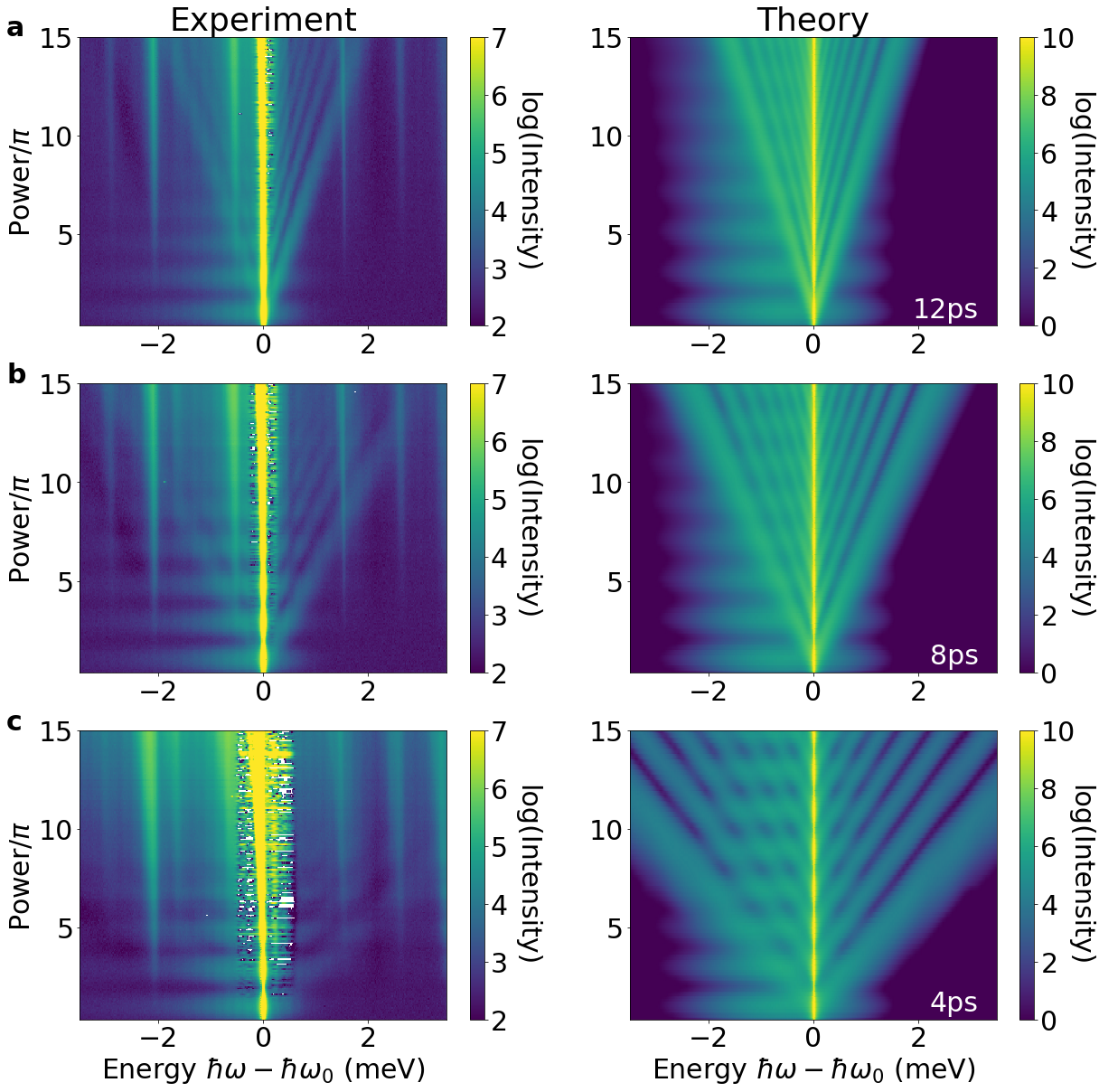}
    \caption{Power-dependent measured and simulated emission spectra from pulsed resonant excitation for different pulses with duration of \textbf{a}, $\SI{12}{\pico \second}$, \textbf{b}, $\SI{8}{\pico \second}$ and \textbf{c}, $\SI{4}{\pico \second}$. With increasing power, Rabi rotations of the main emission peak at zero energy can be observed, while multiple sidepeaks emerge from the center. With shorter excitation pulses the sidepeak spectrum is broader and weaker in intensity.}
    \label{fig:graph2}
\end{figure}

%discussion of graph2
For a pulse with a full-width at half maximum (FWHM) of $\SI{12}{\pico\second}$ with respect to the intensity - which is short compared to the excited state lifetime of $\SI{420}{\pico\second}$~\cite{boos2022} - the power-dependent emission spectra of both experiment (left) and simulation (right) are shown in Fig.~\ref{fig:graph2}a on a logarithmic intensity scale. The excitation power, normalized to the power corresponding to a pulse area of 1$\pi$, is increased from 0 (bottom) to 15 (top). At zero detuning, we observe one prominent emission line corresponding to the initial trion transition energy which, with increasing excitation intensity, exhibits Rabi rotations. As the phonon interaction is non-linear as function of pulse area, thus also excitation-power dependent, in the experiment, excitation-induced re-normalization gives rise to faster oscillation and therefore an offset between the actual oscillation frequency and the normalized pulse area (y-axis)~\cite{krugel2005, Ramsay2010}. For high excitation intensities dephasing damps the oscillation and noise from the suppressed laser pulse overlays this emission. In addition to this modulated center frequency emission line, we observe the characteristic sidepeaks of dynamically dressed states emerging from the center with a periodicity of $2\pi$ and which are lower in intensity by a factor of 5 or more (see supplemental material~\cite{Supplemental}), and their energy shifting linearly away from the laser energy, in agreement with predictions by Moelbjerg \textit{et al.}~\cite{Moelbjerg2012}.
The intensity of the sidepeaks is independent of the Rabi rotations of the main emission, indicating that this part of the emission is generated during the presence of the pulse and is independent of the final state after the entire pulse is absorbed. This leads to the sidepeak emission being dominant for pulse areas of $2n / \pi$, $n\in\mathbb{N}^0$, where the emission from the center peak is weakest (see supplemental material~\cite{Supplemental}). Furthermore, the sidepeaks are still visible for excitation intensities where the Rabi oscillation is strongly damped. \\
In addition, the exciton-phonon interaction results in the well-known phonon sideband~\cite{Besombes2001, Krummheuer2002, favero2003}. In the spectra this is visible as broad horizontal emissions lines  at the red-detuned side of the QD transition in both theory and experiment. The phonon sidebands follow the intensity variation of the Rabi rotation. Additional vertical lines appear in the experimental data at higher excitation power which we attribute to non-resonantly driven states of other QDs in the vicinity of the investigated one.\\
The emission characteristics of the dynamically dressed states strongly depend on the ratio between pulse duration of the excitation and excited-state lifetime, as can be seen when decreasing the pulse duration to $\SI{8}{\pico\second}$ (Fig.~\ref{fig:graph2}b) and $\SI{4}{\pico\second}$ (Fig.~\ref{fig:graph2}c), respectively. The spectral width of the whole sidepeak spectrum, i.e. the detuning of the sidepeaks from the main emission line, increases with shorter and thus spectrally broader pulses (for detailed analysis see supplemental material~\cite{Supplemental}), while the width of the individual sidepeaks increases as well. For a given excitation pulse area, a shorter pulse duration leads to a higher field intensity and thus higher instantaneous Rabi energy, resulting in a stronger dressing of the states which sets the energy of the sidepeaks, while, at the same time, the spectral width of the pulse determines the spectral width of the sidepeaks~\cite{florjaczyk1985}.
Simultaneously, with shorter pulse duration the contribution of the sidepeak to the overall emission decreases (see supplemental material~\cite{Supplemental}). Given that the emission into the sidepeaks occurs during the presence of the pulse, shorter excitation pulses inevitably lead to a shorter presence and variation of the dressed states and thus to less emission into the dressed part of the spectrum. 

\begin{figure}
    \centering
    \includegraphics[width=\linewidth]{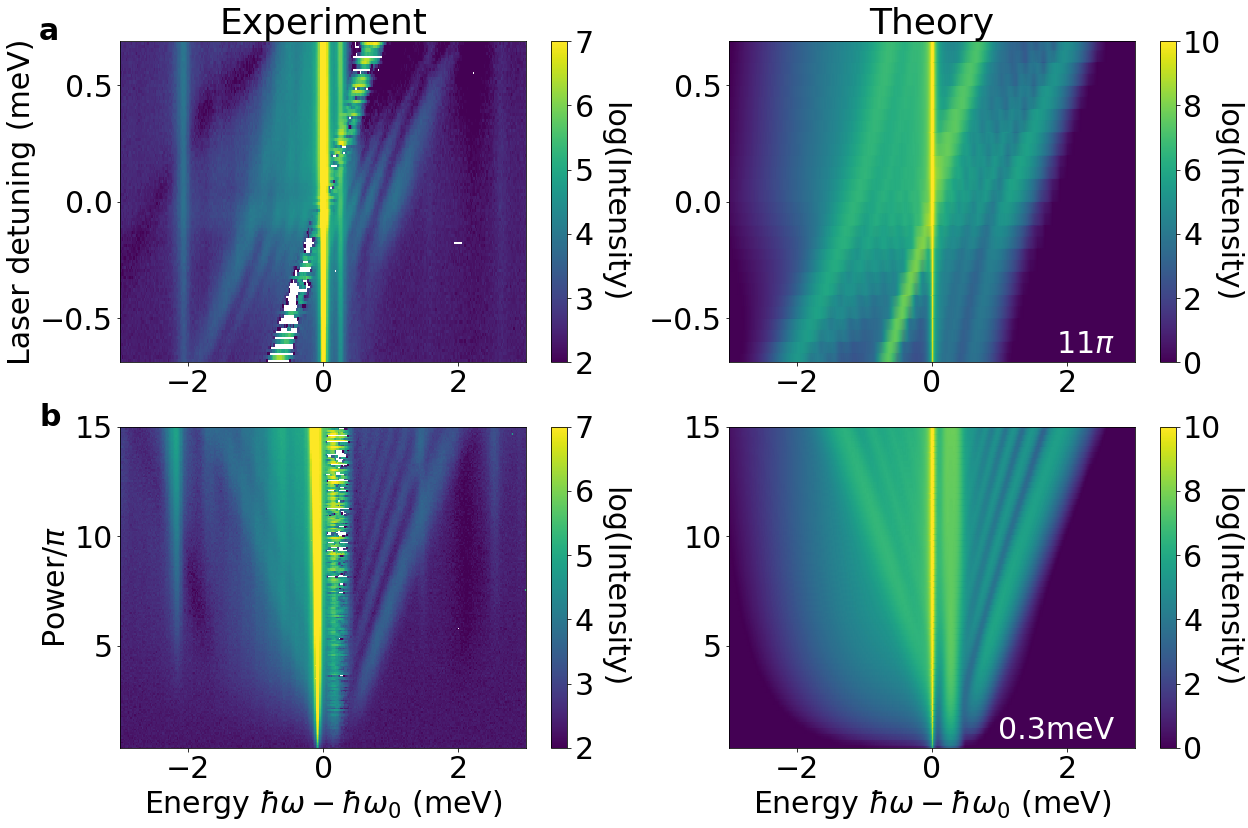}
    \caption{Emission spectra under pulsed excitation with finite detuning. \textbf{a}, The excitation laser ($\SI{12}{\pico \second}$ duration, $11\pi$ pulse area) is detuned in a range of $\pm \SI{0.7}{\milli \eV}$. The spectral detuning of the sidebands in the emission increases with detuning. A prominent peak at the initial frequency of the QD is seen, in addition to a broad spectrum with multiple sidebands centered at the frequency of the laser. \textbf{b}, Power-dependent emission spectra with a $\SI{0.3}{\milli \eV}$ blue-detuned $\SI{12}{\pico \second}$ excitation pulse. The main part of the emission is centered at the energy of the QD, while at the frequency of the laser an additional peak with emerging sidebands can be seen.
    }
    \label{fig:graph34}
\end{figure}

% discussion of graph3
To further study the emission spectrum of dynamically dressed states, we investigate the effect of finite detuning of the excitation pulse (Fig.~\ref{fig:graph34}a). As before, experimental data and simulations are plotted in the left and right panel of the figure, respectively. The $\SI{12}{\pico \second}$ excitation pulse is scanned from $\SI{-0.7}{\milli \eV}$ across resonance to $\SI{+0.7}{\milli \eV}$ detuning, at a constant power corresponding to $11 \pi$. In analogy to emission under continuous-wave excitation, the detuning of the excitation laser introduces an additional splitting of the dressed states~\cite{ulhaq2012, Ulhaq2013}, thus the splitting of the sidepeaks increases with higher absolute detuning, while this whole dressed part of the emission is centered around the laser energy. These characteristics hold true for detuning at other excitation intensities (see supplemental material~\cite{Supplemental}). As expected from the Mollow triplet under continuous-wave excitation, a prominent peak at the energy of the laser is visible in the simulation, with an approximate width corresponding to the width of the laser pulse. Due to the low signal-to-noise ratio for this emission line caused by the noise of the suppressed laser it is only indicated in the experimental data for lower excitation intensities (see supplemental material~\cite{Supplemental}). In analogy to measurements shown above, a bright emission line remains positioned at the energy of the initial QD transition, independent of the energy of the laser and the dressed spectrum.\\ 
To further support these statements experimentally, we look at power-dependent emission under blue-detuned excitation relative to the QD resonance of $\SI{0.3}{\milli \electronvolt}$ (Fig.~\ref{fig:graph34}b). Similar emission spectra as for resonant pulsed excitation can be seen, while now the dressed part of the emission is centered at the position of the laser, consistent with data in Fig.~\ref{fig:graph34}a. Multiple sidepeaks, with increasing splitting for higher pulse power can be observed. At low excitation power where the laser suppression is working best, the predicted peak at the frequency of the laser can be seen, whereas the main part of the emission is centered at the frequency of the QD. Interestingly, even though the sidepeaks seem to originate at the laser energy, they are not directly connected to the central line. \\ 
\\
% summary and acknowledgements and funding
In summary, we have experimentally investigated the resonance fluorescence emission characteristics of a QD excited with finite Gaussian laser pulses. Theoretical calculations are in excellent agreement with the experimental data and corroborate our findings. The resonance fluorescence spectrum reveals multiple sidepeaks that originate from the dynamical dressing of the QD, whereby their number and spectral detuning depends on the excitation intensity as predicted by Moelbjerg \textit{et al.}~\cite{Moelbjerg2012}. Further investigation shows the strong dependence of this characteristic emission spectrum on the excitation pulse length, and on the relative detuning of the excitation laser to the QD resonance. 
Our findings help understand the complex interaction and emission characteristics that arise already in the fundamental setting of a two-level system resonantly driven by a finite pulse. We expect this to be of paramount importance for realizing sources of complex non-classical states of light beyond single-photons and entangled photons~\cite{munoz2014, muller2015, diaz2021, Casalengua2020, Casalengua2023}.\\
\\
We gratefully acknowledge financial support from the German Federal Ministry of Education and Research via the funding program Photonics Research Germany (Contract No. 13N14846), the European Union's Horizon 2020 research and innovation program under Grants Agreement No. 862035 (QLUSTER) and the Deutsche Forschungsgemeinschaft (DFG, German Research Foundation) via projects MU 4215/4-1 (CNLG), INST 95/1220-1 (MQCL) and INST 95/1654-1 FUGG and Germany's Excellence Strategy (MCQST, EXC-2111, 390814868). F.B. gratefully acknowledges the Exploring Quantum Matter (ExQM) program funded by the State of Bavaria. 

\newpage
\onecolumngrid
\newpage

\section{Supplemental Material}

\section{Additional measurements and simulations for resonant excitation and varying pulse length}
\begin{figure}[b]
    \centering
    \includegraphics[width=0.65\linewidth]{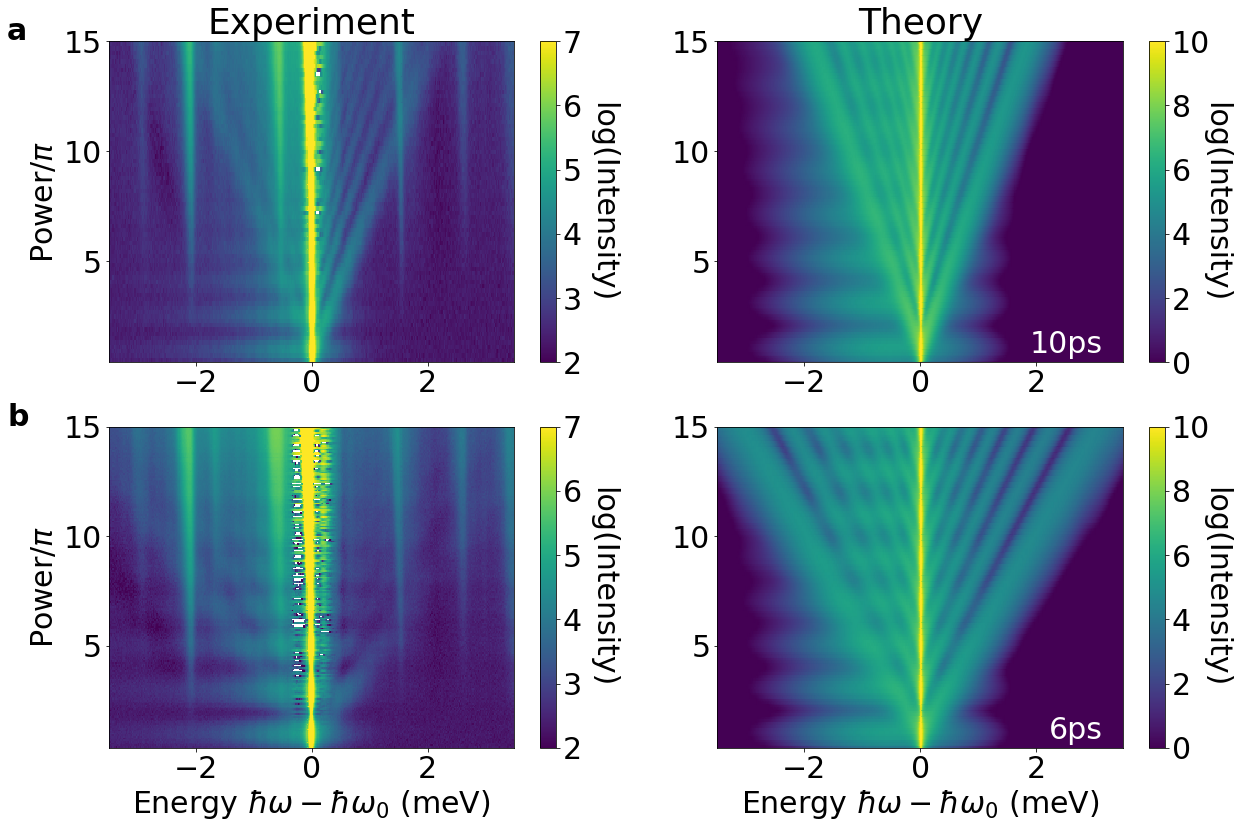}
    \caption{Excitation power dependent emission spectra for resonant excitation with pulse lengths of \textbf{a}, $\SI{10}{\pico\second}$ and \textbf{b}, $\SI{6}{\pico\second}$.}
    
    \label{fig:supp_1}
\end{figure}

Additional power-dependent emission spectra for pulse lengths of $\SI{10}{\pico\second}$ and $\SI{6}{\pico\second}$ are presented in Fig.~\ref{fig:supp_1}a and Fig.~\ref{fig:supp_1}b, respectively. The intensity oscillation in the resonant emission peak confirms coherent driving of the studied two-level system, while the phonon sideband is visible at the lower-energy side of the QD transition, as discussed in the main manuscript.
In analogy to the emission spectra for different pulse lengths shown in the main manuscript, one can unambiguously identify the emission into multiple sidepeaks for increasing excitation power. The emission spectra for pulse lengths of $\SI{10}{\pico\second}$ (Fig.~\ref{fig:supp_1}a) and $\SI{6}{\pico\second}$ fit well into our observations of the pulse-length dependent features of the sidepeaks.

\section{Quantitative analysis of sidepeak position and intensity}
\begin{figure}
    \centering
    \includegraphics[width=0.75\linewidth]{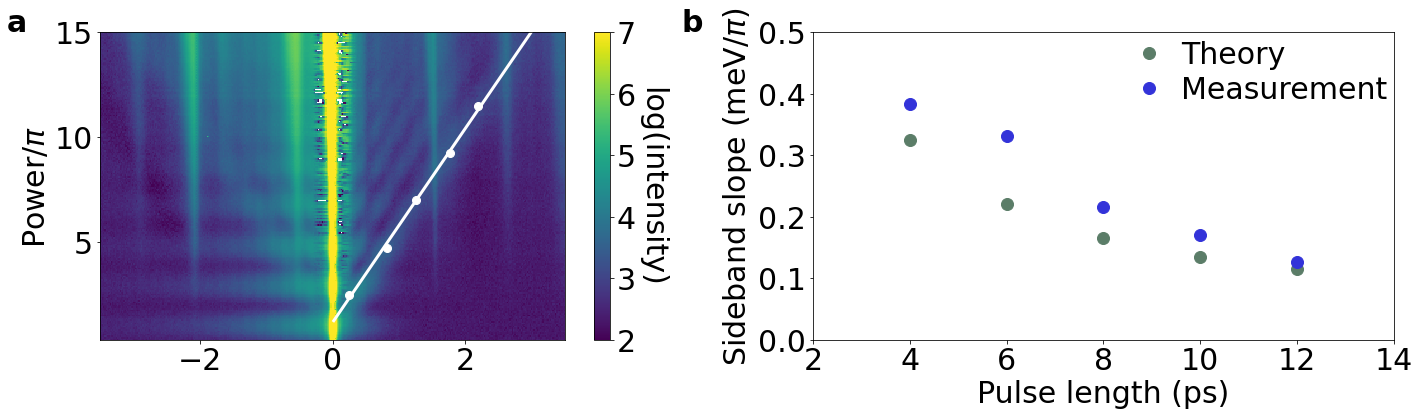}
    \caption{Analysis of the sidepeak slope in experiment and theory. \textbf{a}, Maxima of the sidepeak from five spectra at different pulse powers (white symbol) and their first order polynomial fitting (white solid line) are depicted on power-dependent emission spectra with an exemplary pulse length of $\SI{8}{\pico\second}$. \textbf{b}, Slope of the outermost sidepeak for experimental (blue) and simulated data (green) in dependence of pulse length.}
    \label{fig:supp_2-1}
\end{figure}

To quantify the observation explained in the main manuscript on the characteristics of the sidepeaks depending on the pulse length, we approximate the slope of the outermost sidepeaks linearly. Outermost peak maxima (symbols) in the experimental data with pulse length of $\SI{8}{\pico\second}$ and their linear fit (solid line) are presented in Fig.~\ref{fig:supp_2-1}a. The slope for each pulse length is extracted from five local maxima of the positive-detuned outermost sidepeak, which is linearly fitted afterwards. We depict them as a function of pulse length in Fig.~\ref{fig:supp_2-1}b. For both experimental and theoretical data, a clear decrease in the slope of the sidepeak for longer pulses is visible, from about around $\SI{0.4}{\milli \eV}$/$\pi$ to $\SI{0.1}{\milli \eV}$/$\pi$. We attribute the offset between experiment and theory for shorter pulses to deviation in pulse form and from other experimental uncertainties such as power imprecision. 

\begin{figure}
    \centering
    \includegraphics[width=0.75\linewidth]{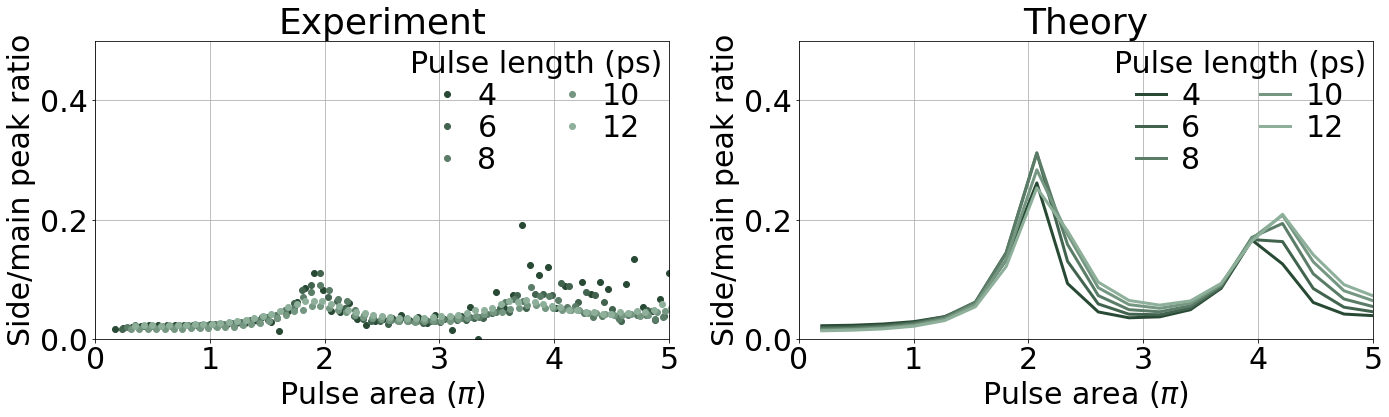}
    \caption{Approximated ratio of sidepeak emission intensity to main peak emission from experiment results and from theoretically calculated emission spectra.}
    \label{fig:supp_2-2}
\end{figure}

Furthermore, we approximate the contribution of emission into the sidepeaks with respect to the center peak emission dependent on pulse area, i.e. excitation power and pulse duration in the measured and theoretically calculated spectra. To this end, we integrate the sidepeaks within a range from $\SI{0.08}{\milli \eV}$ ($\SI{-5.95}{\milli \eV}$) to $\SI{5.95}{\milli \eV}$ ($\SI{-0.08}{\milli \eV}$) for a positive (negative) detuning region and compare it to the center peak integrated within a range of $\SI{-0.08}{\milli \eV}$ to $\SI{0.08}{\milli \eV}$. 
The results are depicted in Fig.~\ref{fig:supp_2-2}. The Rabi rotations of the center peak give rise to the oscillatory behavior of the ratio. An increasing contribution of the sidepeaks to the center peak emission at pulse areas of even $\pi$ further confirms that the emission into the sidepeaks does not follow the Rabi rotation of the central emission line. Note that the dressed part of the emission exhibits a broad peak at the resonant frequency of the QD, which is here partly integrated as a contribution to the center peak. Overall, for longer pulses, the contribution of the sidepeaks to the overall emission increases, as would be expected, as the presence of the pulse and thus the time-dependent dressing of the states increases compared to the fixed lifetime of the excited state. This is less pronounced in the experimental data due to dephasing and noise at the energy of the center peak following the Rabi rotation. 

\section{Additional measurements and simulations for varying detuning at fixed pulse area}
\begin{figure}[h]
    \centering
    \includegraphics[width=0.7\linewidth]{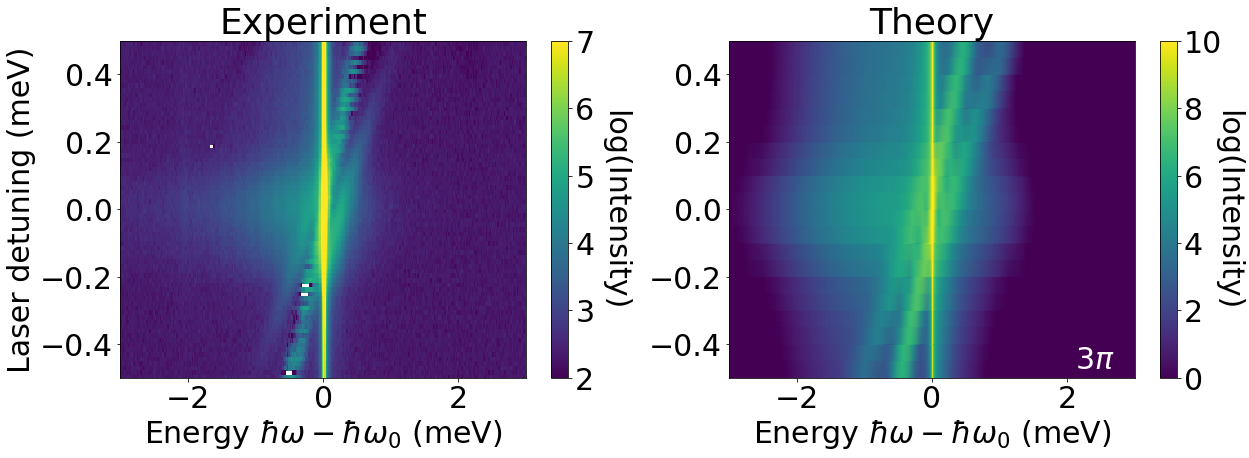}
    \caption{Detuning-dependent emission spectra with a $\SI{12}{\pico\second}$ pulse fixed at pulse area of $3 \pi$.}
    \label{fig:supp_3}
\end{figure}

To verify that the characteristics for the finite detuning of the excitation pulse hold true for different excitation powers, we show experimental and theoretical emission spectra for a $\SI{12}{\pico\second}$ pulse at a pulse area of $3 \pi$, where we scan the detuning within $\pm \SI{0.5}{\milli\electronvolt}$ (Fig.~\ref{fig:supp_3}). 
While multiple sidepeak are clearly resolved for an excitation pulse area of $11 \pi$, only one sidepeak on each side can be observed for a pulse area of $3 \pi$. Overall, a similar behaviour of the emission spectra as for a pulse area of $11 \pi$ can be observed. In addition, in the experimental data the excitation with a pulse area of $3 \pi$ reveals the emission positioned at the excitation energy. This is expected from the simulations, while in the experimental data the emission line is difficult to observe for higher intensities due to limited signal-to-noise ratio at the spectral position of the suppressed laser. 

\section{Theory}
%\noindent{}
Here, we give a brief overview of the most relevant quantities in the theory. The system consists of a ground state $\ket{g}$ and an excited state $\ket{x}$ separated by the energy $\hbar\omega_0$ and is driven by a time dependent term $\Omega(t)$ within the rotating wave approximation (RWA). For a QD, the driving is given in the dipole moment approximation, where $\Omega$ is proportional to the product of the electric field $E(t)$ and the dipole moment with $\Omega(t)=dE(t)/\hbar$. 
The Hamiltonian describing this system reads
\begin{equation}
    H = \hbar\omega_0\ket{x}\bra{x} - \frac{\hbar}{2}\left(\Omega^*\ket{g}\bra{x} + \Omega\ket{x}\bra{g}\right).
\end{equation}
Due to the RWA the driving is given with its complex time dependence with frequency $\omega_L$ with the envelope being a Gaussian with pulse area $\alpha$ and is given by
\begin{equation}
    \Omega(t) = \Omega_0(t) e^{-i\omega_L t} = \frac{\alpha}{\sqrt{2\pi\sigma^2}} e^{-t^2/(2\sigma^2)}e^{-i\omega_L t}
\end{equation}
In the theoretical calculation the pulse duration $\sigma$ corresponds to the electric field, while in the experimental data, the pulse duration $\tau_{p}$ corresponds to the intensity full width of half maximum (FWHM). These quantities are connected via
\begin{equation}
    \tau_p = 2\sqrt{\ln2}\sigma \,.
\end{equation}
To calculate the dynamics of the system we change into a rotating frame. 

We include the photon emission between excited- and ground state as a radiative decay with rate $\gamma$ via Lindblad operators
\begin{equation}
    \mathcal{L}_{\hat{O},\gamma}\rho = \frac{\gamma}{2}\left(2\hat{O}\rho\hat{O}^\dagger - \hat{O}^\dagger\hat{O}\rho-\rho\hat{O}^\dagger\hat{O}\right).
\end{equation}
To allow the transition from $x$ to $g$ we use $\mathcal{L}_{\ket{g}\bra{x},\gamma}$. For the calculations of the time-dynamics we initially assume the system to be in its ground state $\ket{g}$. The decay rate $\gamma$ is set to $1/\gamma = \SI{65}{ps}$, which corresponds to a lifetime shorter than that of the dot used in the experiment (approx. $\SI{420}{\pico\second}$). However, we checked that this difference does not lead to substantial changes in the characteristics of the spectrum, other than longer computation times. \\
The spectrum of the pulsed excitation is calculated via the two-time correlation function $G^{(1)}(t,\tau) = \braket{\sigma^\dagger(t+\tau)\sigma(t)}$, with $\sigma = \ket{g}\bra{x}$. The spectrum is then retrieved taking the Fourier transform with respect to $\tau$ and subsequent integration over $t$,
\begin{equation}
    S(\omega) = \mathrm{Re}\left[\int_{\infty}^{\infty}\!dt\int_{\infty}^{\infty}\!d\tau\,G^{(1)}(t,\tau)e^{-i\omega\tau}\right].
\end{equation}
For the interaction with the phonon environment, we use the standard pure-dephasing Hamiltonian describing the coupling to longitudinal acoustic phonons reading
\begin{align}
\begin{split}
        H_{\text{ph}} &= \hbar\sum_{\q}\omega_{\q}^{}b^{\dagger}_{\q}b_{\q}^{}
        + \hbar\ket{x}\bra{x}\sum_{\q}\left(g_{\q}^{}b_{\q}^{}+g_{\q}^{*}b_{\q}^{\dagger}\right).
\end{split}
\label{eq:hamiltonian_phonons}
\end{align}
Here, $b^{}_{\q}(b^{\dagger}_{\q})$ is a bosonic operator destroying (creating) a phonon with wave vector $\q$ and corresponding energy $\hbar\omega_{\q}$. Coupling between phonons and exciton states is mediated by the coupling element $g_{\q}^{}$. We consider coupling to longitudinal acoustic phonons with linear dispersion and deformation potential coupling. We account for the polaron shift in resonant excitation. The strength of the carrier-phonon interaction is quantified by the phonon spectral density
\begin{equation}
    J(\omega) = \sum_{\q}|g_{\q}|^2\delta(\omega-\omega_{\q}).
\end{equation}

This quantity is highly dependent on the size of the QD, while for the electronic properties only, the specific shape of the dot is not important \cite{luker_phonon_2017}. We assume a $\SI{5}{nm}$ spherical InGaAs QD and take the material parameters from Ref. \cite{kaldewey_coherent_2017} (supplemental material). The resulting phonon spectral density has its maximum of the phonon spectral density at approximately $\SI{1.2}{\milli\electronvolt}$.

\newpage

\providecommand{\noopsort}[1]{}\providecommand{\singleletter}[1]{#1}%

\end{document}